\documentclass{aastex}
\usepackage{emulateapj5}

\shorttitle{HETE Observations of GRB010921 
 }
\shortauthors{Ricker et al.}

\begin{document}
\title{GRB010921: Localization and Observations by the HETE
Satellite {\bf} }

\author{
G. Ricker\altaffilmark{1}
K. Hurley\altaffilmark{6}
D. Lamb\altaffilmark{9}
S. Woosley\altaffilmark{8}
J-L Atteia\altaffilmark{4}
N. Kawai\altaffilmark{2,16}
R. Vanderspek\altaffilmark{1}
G. Crew\altaffilmark{1}
J. Doty\altaffilmark{1}
J. Villasenor\altaffilmark{1}
G. Prigozhin\altaffilmark{1}
G. Monnelly\altaffilmark{1}
N. Butler\altaffilmark{1}
M. Matsuoka\altaffilmark{13,15}
Y. Shirasaki\altaffilmark{14,15,16}
T. Tamagawa\altaffilmark{16}
K. Torii\altaffilmark{16}
T. Sakamoto\altaffilmark{2,16}
A. Yoshida\altaffilmark{13,16}
E. Fenimore\altaffilmark{3}
M. Galassi\altaffilmark{3}
T. Tavenner\altaffilmark{3}
T. Donaghy\altaffilmark{9}
C. Graziani\altaffilmark{9}
M. Boer\altaffilmark{4}
J-P Dezalay\altaffilmark{4}
M. Niel\altaffilmark{4}
J-F Olive\altaffilmark{4}
G.Vedrenne\altaffilmark{4}
T. Cline\altaffilmark{9}
J.G. Jernigan\altaffilmark{6}
A. Levine\altaffilmark{1}
F. Martel\altaffilmark{1}
E. Morgan\altaffilmark{1}
J. Braga\altaffilmark{10}
R. Manchanda\altaffilmark{11}
G. Pizzichini\altaffilmark{5}
K. Takagishi\altaffilmark{12}
M. Yamauchi\altaffilmark{12}
}
\altaffiltext{1}{Massachusetts Institute of Technology, Center for Space
Research, Cambridge, MA, USA.}

\altaffiltext{2}{Department of Physics, Tokyo Institute of Technology, Japan.}

\altaffiltext{3}{Los Alamos National Laboratories, Los Alamos, NM, USA.}

\altaffiltext{4}{Centre d'Etude Spatiale des Rayonnements, Toulouse, France.}

\altaffiltext{5}{Consiglio Nazionale delle Ricerche, istituto TESRE,
Bologna, Italy.}

\altaffiltext{6}{University of California, Berkeley Space Sciences
Laboratory, Berkeley, CA, USA.}

\altaffiltext{7}{University of California, Santa Cruz, CA, USA.}

\altaffiltext{8}{NASA/Goddard Space Flight Center, Greenbelt, MD, USA.}

\altaffiltext{9}{Department of Astronomy and Astrophysics, University
of Chicago, Chicago, IL, USA.}

\altaffiltext{10}{Instituto Nacional de Pesquisas Espaciais, Sao Jose dos
Campos, Brazil.}

\altaffiltext{11}{Tata Institute of Fundamental Research, Mumbai, India.}

\altaffiltext{12}{Faculty of Engineering, Miyazaki University, Japan.}

\altaffiltext{13}{Department of Physics, Aoyama Gakuen University, Tokyo, Japan.}

\altaffiltext{14}{Japan Science and Technology Corporation (JST), Japan.}

\altaffiltext{15}{National Space Development Agency of Japan (NASDA), Japan.}

\altaffiltext{16}{RIKEN (The Institute of Physical and Chemical Research),
Wako, Saitama, Japan.}

\begin{abstract}
On September 21 at 18950.56 SOD (05:15:50.56) UT the FREGATE $\gamma$-ray instrument on
the High Energy Transient Explorer (HETE) detected a bright gamma-ray burst
(GRB).  The burst was also seen by the X-detector on the WXM X-ray instrument and was
therefore well-localized in the X direction; however, the burst was outside the
fully-coded field-of-view of the WXM Y-detector, and therefore information on the Y
direction of the burst was limited. Cross-correlation of the HETE and {\it Ulysses}
time histories yielded an Interplanetary Network (IPN) annulus that crosses the HETE
error strip at a
$\sim$45 degree angle.  The intersection of the HETE error strip and the IPN annulus
produces a diamond-shaped error region for the location of the burst having an area of
310 square arcminutes.  Based on the FREGATE and WXM light curves, the duration of the
burst is characterized by a
$t_{90}$ = 18.4 s in the WXM 4 - 25 keV energy range, and 23.8 s and 21.8 s in the
FREGATE 6 - 40 and 32 - 400 keV energy ranges, respectively.  The fluence of the burst
in these same energy ranges is 4.8 $10^{-6}$, 5.5 $10^{-6}$, and 11.4 $10^{-6}$ erg
cm$^{-2}$, respectively. Subsequent optical and radio observations by ground-based
observers have identified the afterglow of GRB010921 and determined an apparent
redshift of z = 0.450.
\end{abstract}


\section{Introduction}
\setcounter{footnote}{0}

As has long been recognized, accurate locations and
rapid follow-up observations in many wavelengths are central to
understanding the nature of gamma-ray bursts. For this reason,
a strategy evolved in the late 1970's and early 1980's (e.g., Woosley et al. 1982) to
detect GRBs, not only in gamma-rays, but also in emitted X-rays and
optical light. While detection of short transients at the lower energies
posed observational challenges, and the strength of the
optical signal was unknown, the possibility of arc minute
localizations deduced from the X-rays that were known to be present was very
appealing. This strategy was implemented in the HETE-1 (High Energy
Transient Explorer) satellite, which was unfortunately lost due to a
rocket failure on 1996 November 4, and in the highly successful
\textit{BeppoSAX} Mission (Costa et al. 1997).

The HETE-2 satellite (henceforth simply ``HETE''), which was successfully
launched into equatorial orbit on 9 October 2000, is the first space
mission entirely devoted to the study of gamma-ray bursts (GRBs).
HETE utilizes a matched suite of low energy X-ray, medium energy
X-ray, and gamma-ray detectors mounted on a compact spacecraft. A
unique feature of HETE is its capability for localizing GRBs with
$\sim$1-10{\arcmin} accuracy in real time aboard the spacecraft.
GRB locations are transmitted, within seconds to minutes, directly to
a dedicated network of telemetry receivers at 13 automated ``Burst
Alert Stations" (BAS) sited along the satellite ground track
(Villasenor et al 2002). The BAS network then re-distributes the GRB
locations world-wide to all interested observers via Internet and the
GRB Coordinates Network (GCN) in $\approx$1 s (Vanderspek et al. 2002, Barthelmy et
al. 2002).  Thus, prompt optical, IR, and radio follow-up identifications can be
anticipated for a large fraction of HETE GRBs.

Here we report the localization of the first HETE-discovered
\footnote{HETE detects $\sim$50 GRBs yr$^{-1}$, of which $\sim$15 yr$^{-1}$ are
localized by the WXM (Ricker et al. 2002)}
 GRB for which a counterpart has been found. Based on the combined
data from HETE and the Interplanetary Network (IPN), a diamond-shaped error box
roughly 15 arc minutes on an edge was established for GRB010921. Subsequent searches
of this error box by ground-based optical and radio instruments revealed a fading
counterpart whose properties are reported elsewhere (Price et al 2001a, 2002; Kulkarni
et al 2002; Park et al 2001a, 2002; Djorgovski et al. 2001).
  
\section{HETE Mission and Instruments}

The HETE spacecraft is a small satellite, measuring roughly a meter
high by half a meter in diameter, with a mass of 125 kg.  Constructed and 
launched at a cost less than $\frac{1}{3}$ that of a NASA Small Explorer (SMEX), HETE 
was primarily developed and fabricated in-house at MIT by a small scientific 
and engineering team, with major hardware and software contributions from 
international partners in France and Japan (Doty et al. 2002). Contributions to
software  development were also made by scientific partners in the US, 
at the Los Alamos National Laboratory, the University of Chicago, and UC-Berkeley. 
Operation of the HETE satellite and its science instruments, along 
with a dedicated tracking and data telemetry network, is carried out 
by the HETE Science Team itself (Crew et al. 2002). Routine activities are largely
automated,  with oversight and monitoring by members of the Science Team at
MIT and at RIKEN.

The gamma-ray burst detection and localization system on HETE consists of
three complementary instruments: the French Gamma-ray Telescope (FREGATE),
Wide-Field X-Ray Monitor (WXM), and the Soft X-Ray Camera (SXC). The manner
in which the three HETE science instruments  operate cooperatively is
described in Ricker et al. (2002). Since GRB010921 was outside the field-of-view of
the SXC, that instrument will not be discussed in this paper.

{\bf French Gamma-ray Telescope (FREGATE).}
FREGATE consists of four, co-aligned cleaved NaI(Tl) scintillators. Its prime
objectives are the detection and spectroscopy of GRBs and the
monitoring of variable X-ray sources. FREGATE is optimally sensitive in the
6-400 keV energy band, and provides limited ($\approx$ $\pi$ sr) localization
information. The FREGATE instrument was developed by Centre d'Etude Spatiale des
Rayonnements (CESR; Toulouse, France). Additional details regarding FREGATE are given
in Table 4, and in Atteia et al (2002). 

{\bf Wide-Field X-Ray Monitor (WXM).}
The Wide-Field X-ray monitor (WXM) consists of two, crossed one-dimensional coded
aperture cameras. The cameras are designated WXM-X and WXM-Y, respectively, indicating
the spacecraft axis along which each is optimized for GRB localization. Two
position sensitive proportional counters (PSPC) are utilized in each camera. The prime
objectives of the WXM are the detection and spectroscopy of GRBs and the monitoring of
variable X-ray sources in the low energy band. The WXM is optimally sensitive in the
2-25 keV energy band, and typically provides $\le$10{\arcmin} burst localizations.
The WXM was developed by RIKEN(Japan) and the Los Alamos National Laboratory.
Additional details regarding the WXM are given in Table 4, and in Kawai et al (2002)
and Shirasaki et al. (2000).

\section{Observations} \label{observations}

Standard HETE operations consist of a survey for transient gamma-ray and X-ray events
in the 2-3 steradians centered on the antisolar point. Observations take place in that
half of the HETE orbit where the instruments' FOV is clear of the Earth. On September
21 at 18950.56 SOD (05:15:50.56) UT, the FREGATE instrument on HETE detected a bright
($>80 \sigma$) GRB.  In this section we describe the localization and properties of
GRB010921 as determined from HETE and IPN observations of this burst.

\subsection{Localization} \label{localization}

GRB010921 was recorded as trigger event H1761 by the FREGATE instrument on HETE,
and was promptly reported as a GCN Notice at 05:16:08 UT, approximately 17 seconds
after the burst.  The burst was also detected at a SNR=7.1 in the WXM X
detector. The projection of the burst direction was offset 24 degrees from the
WXM X-Z plane, resulting in a good ($\pm$10{\arcmin}) localization in the X direction
and crude limits on Y, thereby localizing the source to a thin, long strip. The
projection of the burst direction was offset 39 degrees from the WXM Y-Z plane,
outside the fully-coded FOV. Thus, the WXM Y camera provided no useful information
regarding the Y location of the burst. Analysis of the Y location required the rapid
development and testing by the HETE Team of  non-standard ground analysis software
that would rely on the limited WXM X camera data, delaying the dissemination of the
two-dimensional localization reported in GCN Circular 1096 until approximately 5.1
hours after the GRB. The initial estimate reported in GCN Circular 1096  was that the
localization strip extended {$\sim$}10$^{\rm o}$ in the long direction, and
20$\arcmin$ in the short direction (Ricker et al 2001a). 
%

Further refinement of the strip dimensions reduced the WXM error region to a 5.2$^{\rm
o}$ by 17$\arcmin$ box, with the 4 corners at coordinates:
\\
\\
$\alpha_{J2000}$ = 22$^{\rm h}$55$^{\rm m}$34.7$^{\rm s}$, $\delta_{J2000}$ = 40$^{\rm
o}$25\arcmin55\arcsec\\
$\alpha_{J2000}$ = 22$^{\rm h}$54$^{\rm m}$05.7$^{\rm s}$, $\delta_{J2000}$ = 40$^{\rm
o}$26\arcmin42\arcsec\\
$\alpha_{J2000}$ = 23$^{\rm h}$04$^{\rm m}$30.4$^{\rm s}$, $\delta_{J2000}$ = 45$^{\rm
o}$23\arcmin35\arcsec\\
$\alpha_{J2000}$ = 23$^{\rm h}$02$^{\rm m}$60.0$^{\rm s}$, $\delta_{J2000}$ = 45$^{\rm
o}$24\arcmin18\arcsec\\

Both \textit{Ulysses} and the gamma-ray burst monitor on \textit{BeppoSAX}
observed GRB010921, and an initial IPN error box was circulated
15.2 h after the burst (Hurley et al. 2001).  Triangulation has now been
performed using the final data, resulting in a \it Ulysses \rm to HETE-FREGATE annulus
centered at $\alpha_{J2000}$, $\delta_{J2000}$ = 15$^{\rm h}$29$^{\rm m}$23.6$^{\rm s}$,
67$^{\rm o}$36\arcmin14\arcsec, with radius 60.082 $\pm$ 0.118 $^{\rm o}$ 
(3 $\sigma$).  This annulus intersects the 3 $\sigma$ WXM error box to give a
310 square arcminute error region whose coordinates are given in Table 1; 
the diamond-shaped error region is shown in figure 1. This error box is
larger than reported in Hurley et al. (2001) due to the fact that the initial WXM error
box was quoted for 90\% confidence (ie 2$\sigma$). Both the Ricker et al. (2001a)
2$\sigma$ error region and the 3$\sigma$ error region given here are consistent with
the position of the optical transient discovered by Price et al. (2001a), with the OT
lying 3.6{\arcmin} from the centerline of the WXM error strip, and 3.1{\arcmin}
from the centerline of the IPN error annulus.

\subsection{Time History, Peak Flux and Fluence} \label{history}

The time history of GRB010921 observed by the HETE science instruments is shown in figure
2. The six panels in figure 2 illustrate the energy dependence of the  burst profile
and duration. For FREGATE, all 4 detectors were summed; for the WXM, only the signal
from one illuminated anode (out of six total) in the X-camera was utilized, so as to
maximize the signal-to-noise ratio, since GRB010921 was quite far (43 degrees) off-axis
from the instrument boresight.  Quantitative measures of the burst temporal properties
(i.e., $t_{50}$ and $t_{90}$; Paciesas et al. 1999) are given in Table 2. At the
highest energies measured by FREGATE (32-400 keV band), the GRB010921 light curve
exhibits a high degree of symmetry, with $t_{50}$=7.4s and $t_{90}$=21.8s. At lower
energies, the
$t_{50}$ duration becomes somewhat larger, increasing by
$\sim30\%$; on the other hand, $t_{90}$ is largely unchanged over the 4-400 keV band. 
Overall, the light curve for GRB010921 exhibits a hard-to-soft spectral evolution 
that is common in many GRBs.

The prompt emission by GRB010921 was measured over more than 2 decades in energy by
HETE (Table 3). It is notable that GRB010921's peak energy flux of 2.0 $10^{-7}$ erg
cm$^{-2}$ s$^{-1}$) in the 4-25 keV X-ray band was $\sim$ $\frac{1}{3}$
the peak energy flux in the ``traditional" 50-300 keV gamma-ray band. A
comparison of the peak photon flux in the two bands is even more striking: $\sim$4
times as many 4-25 keV photons were emitted as were 50-300 keV photons. (For
GRB010921, the smaller SNR in the WXM compared to the
SNR in FREGATE arises from the fact that GRB010921 was offset 43$^{\rm
o}$ from the boresight of the WXM.) 

\section{Discussion}

The temporal characteristics of GRB010921 clearly mark it as belonging to the class of
long duration GRBs (Hurley 1992, Kouveliotou et al. 1993). It is also
quite ``X-ray rich," in that it exhibits a value of
$L_{x(4-25keV)}$/$L_{\gamma(50-300keV)}$
$\sim$0.3. However, it is by no means deficient in its 50-300 keV gamma-ray
flux: in fact, its peak flux of 3.6 ph cm$^{-2}$ s$^{-1}$ in the 50-300 keV range
would place it within the top decile of bursts in the 4Br Catalog (Paciesias et al
1999). Thus, GRB010921 does not qualify as an X-ray Flash (XRF; Heise et al. 2001).
Based on its apparent redshift of z= 0.450 (Djorgovski et al. 2001), the
implied isotropic emitted energy of GRB010921 in the 8-400 keV band is 7.8 10$^{51}$
ergs (assuming $\Omega_{M}$=0.3, $\Omega_{\Lambda}$=0.7, $H_{o}$=65 km s$^{-1}$
Mpc$^{-1}$; for beamed emission, the total energy should be multiplied by the
as-yet-unknown beaming fraction.)  The redshift of GRB010921, the second lowest among
established GRBs, makes it a strong candidate for extended searches for a
possible associated supernova. Furthermore, GRB010921 was, like
all HETE-discovered GRBs, transiting near midnight at the initial epoch of the burst,
resulting in its being well-situated for followup studies by ground-based telescopes
for several months following the burst. Thus, additional insight into the nature of
GRB010921 can be anticipated from long-term optical photometry by large telescopes.   

\section{Conclusions}

The localization of GRB010921 by HETE and the IPN  has led directly to a successful
identification of a low-redshift afterglow counterpart. The prompt initial
GCN alert for GRB010921 was distributed worldwide within 17 s of the burst; however, 
the determination of the HETE WXM localization was delayed by 5.1 hours due to the
need to prototype new software to accommodate GRB010921's extreme off-axis position in
one of the WXM cameras. A confirmation, and definitive refinement of the WXM
localization, using IPN data available 15.2 hours after the GRB enabled ground-based
observers to successfully target GRB010921 during the first night following the burst,
and to establish a counterpart well within the HETE-IPN error region. Hopefully, over
the coming months the identification of optical counterparts resulting from prompt
($\le$100 s), accurate HETE localizations will become routine. With the
currently-projected long orbital lifetime ($>$10 years) and excellent health of the
HETE spacecraft and instruments, we look forward to providing a uniquely valuable
service to the worldwide community of GRB observers. 

\acknowledgments
\section*{Acknowledgments}

The combined efforts over almost two decades by many talented and dedicated
individuals made the HETE mission possible. In particular, the efforts of
Bob Dill, Tye Brady, Dave Breslau, Gus Comeyne, Francis Cotin, Jim Francis,	Greg
Huffman, Frank LaRosa, Fred Miller, Jerry Roberts, and Fuyuki Tokanai were particularly
notable.

The HETE mission is supported in the USA by NASA Contract NASW-4690; in Japan, in part
by the Ministry of Education, Culture, Sports, Science, and Technology Grant-in-Aid
13440063. KH is grateful for
\it Ulysses
\rm support under Contract JPL 958059, and for HETE support under Contract
MIT-SC-R-293291. GP acknowledges support by the Italian Space Agency (ASI).

\clearpage

\begin{deluxetable}{cc}
\tablecaption{GRB010921 Error box coordinates.\label{tbl-1}}
\tablewidth{0pt}
\tablehead{
\colhead{$\alpha_{J2000}$} & \colhead{$\delta_{J2000}$}  
}
\startdata

22$^{\rm h}$ 56$^{\rm m}$ 6.9$^{\rm s}$ &   40$^{\rm o}$ 45{\arcmin}
  21{\arcsec}  \\
22$^{\rm h}$ 56$^{\rm m}$ 37.3$^{\rm s}$  &  41$^{\rm o}$ 03{\arcmin}
  28{\arcsec}  \\
22$^{\rm h}$ 54$^{\rm m}$ 21.74$^{\rm s}$  &  40$^{\rm o}$ 36{\arcmin}
  25{\arcsec}  \\
22$^{\rm h}$ 54$^{\rm m}$ 51.9$^{\rm s}$  &  40$^{\rm o}$ 54{\arcmin}
  32{\arcsec}  \\

\enddata
\end{deluxetable}

\begin{deluxetable}{cccr}
\tablecaption{Temporal Properties of GRB010921.\label{tbl-1}}
\tablewidth{0pt}
\tablehead{ \colhead{Energy Band (keV)} &
\colhead{$t_{50}$ (s)}& \colhead{$t_{90}$ (s)}  }

\startdata
4 - 10 & 8.8 & 18.8  \\
10 - 25 & 8.2 & 15.7 \\
32 - 400 & 7.4 & 21.8 \\

\enddata
\end{deluxetable}

\begin{deluxetable}{cccccr}
\tablecaption{Energy Emission Properties of GRB010921.\label{tbl-1}}
\tablewidth{0pt}
\tablehead{Energy Band & Signal-to-Noise
& Peak Photon Flux & Peak Energy Flux & Fluence \\ (keV) & (in 3 measured bands) &
(ph cm$^{-2}$ s$^{-1}$) & (erg cm$^{-2}$ s$^{-1}$) & (erg cm$^{-2}$) }

\startdata
4 - 25 & 7.1  & 14.8 & 2.0 $10^{-7}$ & 2.0 $10^{-6}$ \\
8 - 400 & 87.9 & 17.2 & 1.1 $10^{-6}$ & 15.4 $10^{-6}$ \\
32 - 400 & 43.6 & 6.0 & 8.5 $10^{-7}$ & 11.4 $10^{-6}$ \\
25 - 100 & ---  & 6.1 & 4.7 $10^{-7}$ & 6.2 $10^{-6}$ \\
50 - 300 & --- & 3.6 & 6.1 $10^{-7}$ & 8.4 $10^{-6}$ \\

\enddata
\end{deluxetable}

\begin{figure}
\includegraphics*{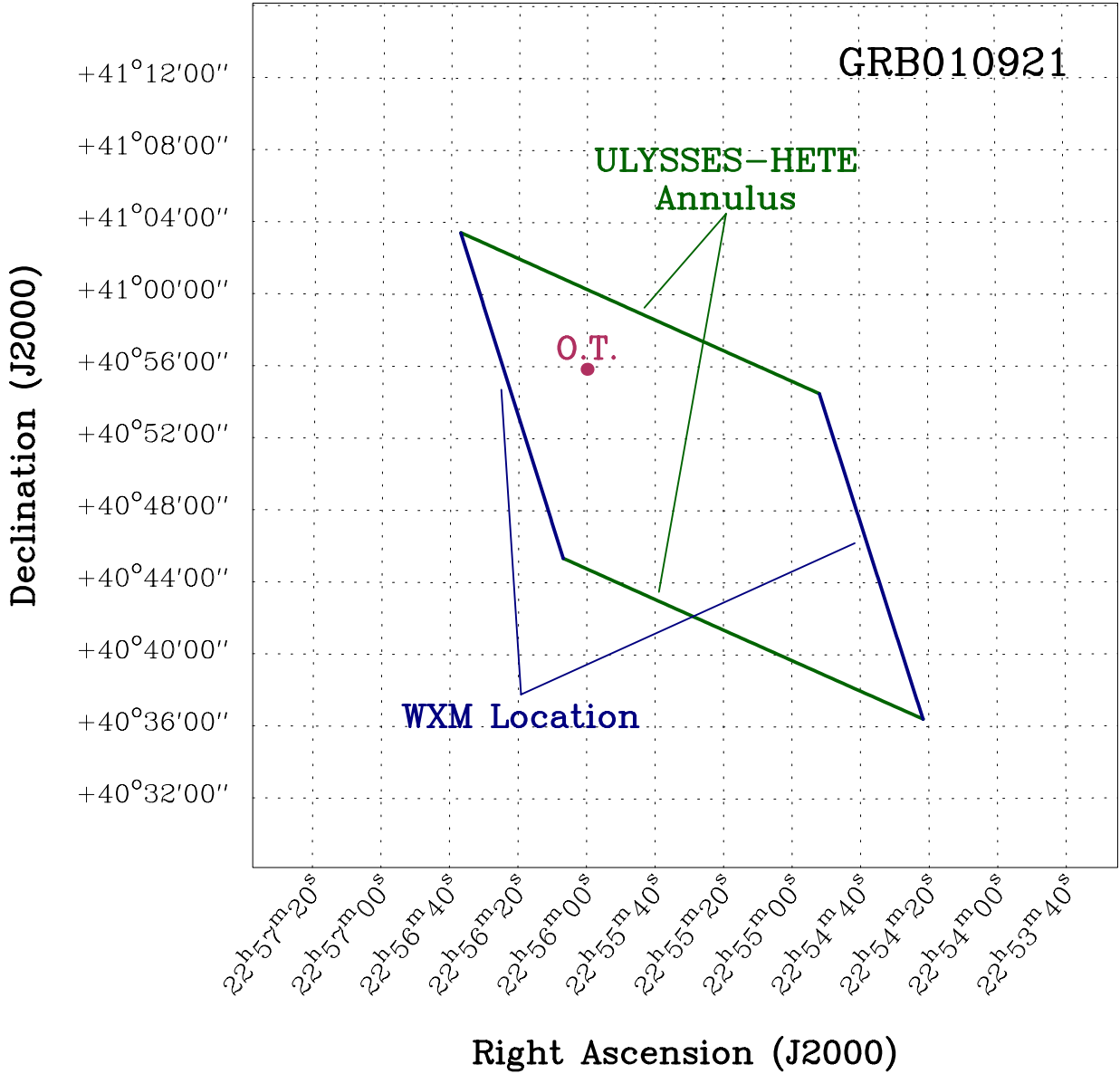}
\caption{The final error box of GRB010921, determined from
superposing the 3 $\sigma$ WXM
error box (north-south trending lines) and the 3 $\sigma$ IPN annulus
(east-west trending lines).  OT indicates the position of the optical
transient source found by Price et al. (2001a), at $\alpha_{J2000}$
 = 22$^{\rm h}$ 55$^{\rm m}$ 59.9$^{\rm s}$, $\delta_{J2000}$ =
40$^{\rm o}$ 55{\arcmin}  53{\arcsec}. 
\label{Fig. 1}}
\end{figure}

\begin{figure}
\includegraphics*{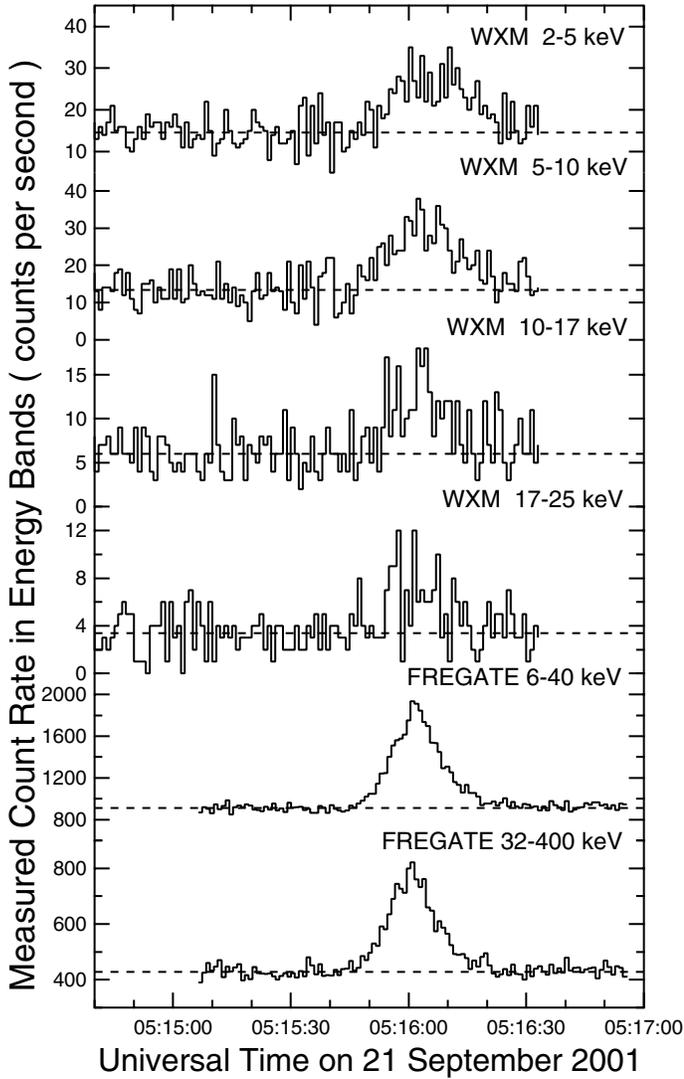}
\caption{WXM and FREGATE light curves for GRB010921. The six panels correspond to
six energy bands; the dashed lines indicate the backgound levels. For the WXM and
FREGATE, the projected effective areas to GRB010921 were 10 cm$^{2}$ and 70
cm$^{2}$, respectively.\label{Fig. 2}}
\end{figure}

\begin{deluxetable}{lll}
\label{table4}
\tablecaption{HETE Instrument Characteristics}
\tablewidth{0pt}
\tablehead{Characteristic & Fregate & WXM}

\startdata
Instrument type&Cleaved Nal(TI) &
  Coded Mask with PSPC \\
Energy Range&6 keV to $>$ 400 keV & 2 to 25 keV\\
Timing Resolution &10 $\mu$s & 1 $\mu$s \\
Spectral Resolution&$\sim$25\% $@$ 20keV, $\sim$ 9\% $@$ 662 keV &
$\sim$ 20\% $@$ 8 keV\\
Effective Area (on axis)& 160 cm$^2$ & $\sim$116 cm$^2$ \\
Sensitivity (10 sigma)& $\sim1$ $10^{-7}$ erg cm$^{-2}$ s$^{-1}$
(50--300 keV)  & $\sim8$ {$10^{-9}$} erg cm$^{-2}$ s$^{-1}$ (2-10 keV)\\
Field of View& $\sim$4 steradians &  $\sim$1.5 steradians \\
\enddata
\end{deluxetable}


\begin{thebibliography}{}

\renewcommand{\thefootnote}{\fnsymbol{footnote}}

\bibitem[Atteia(2002)]{WH2001_FREGATE}
Atteia, J-L, et al. 2002, {\it In-flight Performance and First Results from the FREGATE
Instrument on HETE}, in WH2001\footnote[7]{WH2001 = {\it Gamma-Ray Burst and Afterglow
Astronomy 2001: A Workshop Celebrating the First Year of the HETE Mission}, Woods
Hole, MA, November 2001, to be published in the AIP Conference Proceedings (AIP Press:
New York).}.

\renewcommand{\thefootnote}{\arabic{footnote}}

\bibitem[Barthelmy(2002)]{WH2001_GCN}
Barthelmy, S. D., et al. 2002, {\it GCN: A Status Report}, in WH2001{$^{**}$}.

\bibitem[Bloom(2001)]{gcn1135_HST}
Bloom, J.S., et al. 2001, GCN Circ. 1135.

\bibitem[Costa(1997)]{SAX Ref}
Costa, E., Frontera, F., Heise, J., et al. 1997, Nature, \textbf{387}, 783.

\bibitem[Crew(2002)]{WH2001_MOPS}
Crew, G.B., et al. 2002, {\it HETE Mission Operations}, in WH2001{$^{**}$}.

\bibitem[Djorgovski(2001)]{gcn1108}
Djorgovski, S.G., et al. 2001, GCN Circ. 1108.

\bibitem[Heise(2001)]{XRF_Rome2000}
Heise, J., in't Zand, J., Kippen, R., and Woods, P., {\it X-Ray Flashes and X-Ray Rich
Gamma-Ray Bursts}, in Gamma-Ray Bursts in the Afterglow Era, (Rome, Italy, 17-20
October 2000), ESO Astrophysics Symposia, Springer (Berlin ), p. 16, 2001.

\bibitem[Hurley(1992)]{Short_Long_GRBs_1992}
Hurley, K. 1992, {\it Gamma-Ray Burst Observations: Past and Future}, in Gamma-Ray
Bursts, Eds. W. Paciesas and G. Fishman, AIP Conf. Proc. 265 (AIP Press- New York), 3.

\bibitem[Hurley(2001)]{gcn1097}
Hurley, K., et al. 2001, GCN Circ. 1097.

\bibitem[Kawai(2002)]{WH2001_WXM}
Kawai, N., et al. 2002, {\it In-orbit Performance of the WXM Instrument on HETE}, in
WH2001{$^{**}$}.

\bibitem[Kouveliotou(1993)]{Short_Long_GRBs_1993}
Kouveliotou, C., et al. 1993, Ap. J. Lett., \textbf{413}, L101.

\bibitem[Kulkarni(2002)]{HST_ApJLett}
Kulkarni, S., et al. 2002, in preparation.

\bibitem[Paciesas(1999)]{4Br}
Paciesas, W.S. et al. 1999, ApJS, \textbf{122}, 465P.

\bibitem[Park(2001a)]{gcn1114}
Park, H. S., et al. 2001a, GCN Circ. 1114.

\bibitem[Park(2002)]{LOTIS_ApJLett}
Park, H. S., et al. 2002, submitted to Ap. J. Letters (astro-pf/0112397).

\bibitem[Price(2001a)]{gcn1107}
Price, P.A., et al. 2001a, GCN Circ. 1107.

\bibitem[Price(2002)]{Palomar_ApJLett}
Price, P.A., et al. 2002, submitted to Ap. J. Letters (astro-pf/0201399).

\bibitem[Ricker(2001a)]{gcn1096}
Ricker, G.R., et al. 2001a, GCN Circ. 1096.

\bibitem[Ricker(2002)]{WH2001_HETE}
Ricker, G.R., et al. 2002, {\it High Energy Transient Explorer (HETE): Mission and
Science Overview}, in WH2001{$^{**}$}.

\bibitem[Shirasaki(2000)]{Shirasaki_SPIE}
Shirasaki, Y., et al. 2000, Proc. SPIE, \textbf{4012}, pp 166-177.

\bibitem[Vanderspek(2002)]{WH2001_GCN}
Vanderspek, R.K., et al. 2002, {\it What HETE sends to the GCN}, in WH2001{$^{**}$}.

\bibitem[Villasenor(2002)]{WH2001_BAS}
Villasenor, J.N., et al. 2002, {\it First Year of Operations of the HETE Burst Alert
Network}, in WH2001{$^{**}$}.

\bibitem[Woosley(1982)]{Santa_Cruz_1982}
Woosley, S. E.  et al. 1984, The High Energy Transient Explorer (HETE), in {\it High
Energy Transients in Astrophysics} (Santa Cruz, CA 1983), ed. S. E. Woosley, A.I.P.
Conf. Proc. 115 (AIP Press-New York), p. 709.

\end{thebibliography}
\end{document}